# The Concept, Principles of Design and Implementation of the University Cloud-based Learning and Research Environment

Olena Glazunova[1], Mariya Shyshkina[2],

[1]National University of Life and Environmental Sciences of Ukraine,
15 Heroyiv Oborony st., Kyiv, Ukraine
`o-glazunova@nubip.edu.ua`

[2]Institute of Information Technologies and Learning Tools of NAES of Ukraine,
9 M.Berlynskoho St., Kyiv, Ukraine
`shyshkina@iitlt.gov.ua`

**Abstract.** The scientific and methodological background of creation and development of the university cloud-based learning and research environment is substantiated. The conceptual and terminology body of the cloud-based environment investigation is defined, the main features of such environment are revealed. The main methodological principles of the environment design and development are considered among them there are the principles of open education, open science and also the specific principles inherent to the cloud-based systems. The general model of the university cloud-based learning and research environment formation is substantiated and six main stages of this environment formation are distinguished in the model. The cloud-based environment functions, content and tools are revealed in accordance with the proposed methodological principles, the criteria for the estimation of this environment efficiency are elaborated. The results of implementation and experimental research of the cloud-based environment formation at the National University of Life and Environmental Sciences of Ukraine are described. The influence of different environment components use on students' success is explored.

## 1 Introduction

### 1.1 The Problem Statement

Nowadays, an effective and promising approach to open pedagogical systems design is the use of the cloud computing technologies to provide ICT support for the functioning and development of the computer-based educational and research environment. Innovative technological solutions for learning environment organization and design using cloud computing (CC) and ICT outsourcing have shown promise and usefulness [2-5]. The challenges of making the ICT infrastructure of the university environment fit the needs of its users, taking maximum advantage of modern network technologies, and ensuring the best pedagogical outcomes, have led to the search for the most reasonable ways of its modernization. So, the modelling and analysis of the processes of cloud-based learning environment formation, design and deployment in view of the current tendencies of ICT advance have come to the fore.

Inconsistence of the structure and composition of the universities learning and research environment with the requirements of constructing pedagogical systems (PS), which are required in view of the modern challenges of higher education development and urgent need of educational practice, insufficient study of theoretical and methodological foundations of the university cloud-based environment development, its information resources and tools are constrains of the university pedagogical systems development hindering further improvement of the quality of higher education.

### 1.2 The Theoretical Background

According to the recent research [3], [4], [8]; [11], [16], [20], [21], [22] the problems of cloud technologies implementing in higher educational institutions to provide software access, support collaborative learning, research and educational activities, exchange experience and also project development are especially challenging. The formation of the cloud-based learning environment is recognized as a priority by the international educational community [14] and is now being intensively developed in different areas of education [9], [10], [13], [17], [20].

Among the current issues there are those concerning existing approaches and models for electronic educational resources delivery within the cloud-based setting; the methodology of CC-based learning and research university environment design; evaluation of current experience of cloud-based models and components use [3], [4], [10], [15], [16], [18], [22]. This brings the problems of the modelling of the cloud-based learning environment structure and development to the fore front. It is important to define the role of different types of information analytical cloud-based learning and research tools and other network electronic educational resources (EER) of open PS, to determine the indicators for measuring the pedagogical effect of certain innovations introduction.

In Ukraine some steps were carried out according to this perspective direction, since the annual international seminar "Cloud Technologies in Education" was initiated in (2012) [15], the activities of joint research laboratories (Institute of Information Technologies and Learning Tools of NAES of Ukraine, Kryvyi Rig National University, Ternopil National Pedagogical University named after Volodymyr Hnatyuk, I.Franko Zhytomyr State University, I.Franko Drohobych Pedagogical University, National University of Life and Environmental Sciences of Ukraine [11].

Theoretical results and practical orientation of the researches conducted at the Institute of Information Technologies and Learning Tools of the NAES of Ukraine are mainly subordinated to the specified educational paradigm, aimed at the development of scientific and methodological foundation of implementing the principles of open education. In particular, in V. Bykov's work "The models of organizational systems of open education" the models of information educational environment were pro-posed. The named works are aimed to create the methodological basis for further researches in this area, given that the cloud-based environment is a new step in the development of open education systems.

In 2015-2017 the research project "Methodology of the formation of the cloud-based learning and research environment of a higher educational institution" was conducted at the Institute of Information Technologies and Learning Tools of the National Academy of Educational Sciences of Ukraine, State Registration №0115U002231, the project coordinator - M. Shyshkina. In the course of this project the conceptual basis, principles and approaches to the environment formation, classification of services, design of the general model of its formation and development were considered and implemented [17].

In recent years a series of dissertations on the application of different types of cloud-based services at higher educational institutions have been completed in Ukraine (O.Glazunova (2015), M.Shyshkina (2016), M.Kyslova (2015), M.Popel (2017), O.Merzlikin (2017) and others http://iitlt.gov.ua/eng/atestat/spetsializovana-vchena-rada/avtoreferaty-dysertatsiyi.php).

### 1.3 The Purpose of the Article

The main purpose of the article is to define the conceptual body and principles of the cloud-based learning environment formation and to consider the possible ways and techniques of its use and application within the open pedagogical systems of higher education. The main idea lies in the hypothesis that design and development of learning and research environment based on the proposed approach will result in positive effect, namely, increase of educational process quality.

### 1.4 The Research Methods

The research methods involved analysing the current research (including the domestic and foreign experience of the application of cloud-based learning services to define the concept of the investigation and research indicators), examining existing models and approaches, technological solutions and psychological and pedagogical assumptions about better ways of introducing innovative technology so as to consider and elaborate the general model of the environment formation and special methodical system of training with the use of the cloud-based components. To measure the efficiency of the proposed approach the pedagogical experiment was undertaken. The special indicators to reveal the efficiency of the cloud-based learning environment were proposed and substantiated.

## 2 Discussion

The use of ICT affects the content, methods and organizational forms of learning and managing educational and research activities that require new approaches to learning environment arrangement [2], [3]. Therefore, the formation of modern cloud-based systems for supporting learning and research activities should be based on appropriate innovative models and methodology that can ensure a harmonious combination and embedding of various networking tools into the educational environment of higher education institution [2], [3], [5], [11], [12], [19].

The cloud-based learning and research environment (CBLRE) of a higher educational institution is the environment in which the virtualized computer-technological (corporate or hybrid-based) infrastructure is purposefully built for the realization of computer-procedural functions (such as content-technological and information-communication functions) [17].

Essential features of formation and development of the cloud-based learning and research environment of higher educational institutions are such properties as openness and flexibility.

The openness of the cloud-based environment relates to its permanent dynamic relationship with the external to this environment socio-economic space, which sets goals and objectives and defines the requirements for the functioning and development of educational systems, provides them with the necessary resources and utilizes end products. If the resources necessary for functioning and development of educational systems are available, just the openness is the imperative for the systematic adaptation of the environment structure to the tasks and requirements imposed by the external socio-economic space. The ability of the environment to be adaptive, to provide progressive changes is determined by the flexibility of its structure and configuration.

Flexibility of the cloud-based environment as for providing conditions for the development of the target and methodological learning subsystems of a particular pedagogical system means the potential suitability of the environment to changes (in certain predetermined allowable limits) of the composition, structure and parameters of its components, which do not lead to loss of its integrity (the destruction of its system-forming relations, going beyond the intended variability of component parameters), significant changes in its main target and functional subsystems or loss of safety.

The development of the cloud-based environment is an evolutionary change in the structure of the environment and / or its component parameters (for example, the procedural capacity and volumes of storage clusters, the coverage area of access to computer networks through wireless communication channels) that occurs in accordance with the updated goals of the pedagogical system, desirable (planned, predicted, hypothetical) characteristics of its end products at certain stages of development.

This is due to such properties of the cloud-oriented environment as the openness and flexibility of its structure its composition so, that it can be brought into conformity with the planned development goals and new tasks which have arisen or will arise in the near future. These properties potentially enable changes in the tasks of the formation and development of the environment and, as a result, adequate changes in the composition and parameters of its facilities and the modernization of the methods of its design and use.

Flexibility and openness of the environment are achieved through the cloud technologies application. After all, these technologies, the cloud platforms, were originated from the very beginning in view of the reasons of maintaining flexible and open systems. That is why this type of platforms is the most promising to design the computer-based infrastructure for the whole educational institution, as it will be possible to create the best conditions for the progressive development of the environment.

At the same time, the cloud-based environment of an educational institution is a complex system that contains a significant number of subsystems, implements various functions that are formed at the level of the institution, its separate structural subdivision. Special attention is required to the methods of designing and using environment components for different levels of its organization in the implementation of various types of the cloud-based tools. Therefore, a set of techniques may be needed to deploy and use the cloud-based environment or its components.

### 2.1 The Principles of Cloud-based Learning and Research Environment Formation

Among the whole set of psychological and pedagogical principles of the cloud-based environment formation the special attention should be paid to the principles of open education and open science, which are realized in greater extent through the tools of this environment; and the specific principles typical for the cloud-based educational and research systems.

The present research is based on the following principles of open education [1]:

- the principle of *mobility of students and teachers* ensuring mobility of graduates of an education system and teachers on markets (including international) of labour and educational services;
- the principle of *equal access to educational systems* ensuring equal access to education for everyone who has the desire and need to study throughout life and have the opportunity to do so;
- the principle of *qualitative education* concerns to the provision through open systems of such quality of education that would correspond to the individual educational needs of students and the requirements of society regarding the general and professional educational level of its members;
- the principle of *the structure and implementation of educational services formation* is connected to the provision of the market mechanisms for the formation of qualitative and quantitative structure of training, retraining and advanced training of learners, and the implementation of a range of educational services offered through open education systems.

Open science means a radical change in the processes of nature transformation, science and innovation through the integration of ICT into the process of research and the development of online culture of openness and exchange. It is more open, more global and common, more creative, and closer to society [6, 12]. It relies on the use of e-infrastructure, that is, on ICT-based services and tools for data processing and research in virtual and shared environments [7, 22]. On these provisions the principles of open science were defined:

- the principle of *technological development*, which involves the use of methods and tools based on research network infrastructures for powerful distributed computing and data storage to process large amounts of data, conducting of virtual experiments in various fields of science, use the tools and models of the results analysis, interpretation and validation;
- the principle of *open access* to research results and processes that are based on public availability of the results of their research, access to research for other scientists, opportunities for cooperation and data sharing, collaboration in virtual research communities, informal means of cooperation;
- the principle of *research collaboration* involving the use of platforms and infrastructures that support collaborative multidisciplinary research, new levels of scientific collaboration in various fields of science;
- the principle of *interaction with society*, which is characterized by the involvement of citizens and society in participating in the conduct and discussion of the effects of scientific research, the creation of new relations between science and society;
- the principle of *innovative nature* of open science, resulting from understanding the global current problem, such as ecological, energy, demographic, arms, health, and others - and developing appropriate measures for their solution.

To the *specific principles* (typical for the cloud-based systems) we refer the following:

- the principle of *adaptability* concerns to the suitability of learning tools and services of the system for the needs of the broadest possible contingent of users, which may have different information and procedural needs associated with different levels of knowledge, individual characteristics, rate of mastering material, etc.;
- the principle of *personalization* of services delivery ensures a person-oriented (personified) approach to learning by adjusting the ICT infrastructure of the environment (including virtual) to individual information and communication, resource, operational and procedural needs of participants of the educational process;
- the principle of learning and research environment of *ICT infrastructure management unification* presupposes the uniformity of its structure, aimed at integrated data storage and management of large arrays on the unified basis, which is necessary to ensure systematic, invariant approaches to the organization of access to learning and research resources within this environment;
- the principle of full-scale interactivity of learning and research tools of the cloud-based environment relates to the organization of the effective feedback of these tools use and to the support of the interactive mode of cooperation with mobile participants. Through the feedback the control and evaluation of the learner's actions is provided, the permanent access to the guidance and support materials is provided. It is assumed that the feedback really comes up as instant, something that helps in real time, allowing the most comprehensive respond to the needs of the learner;
- the principle of *flexibility and scalability* of access to resources and tools of the cloud-based environment is aimed at dynamic receiving, deploying and supplying of the cloud services and platforms, as well as promptly releasing the computational resources that are not at need increasing the effectiveness of the learning process organization, providing the ability to quickly adapt to the changing requirements and arising problems;
- the principle of *data and resources consolidation* is implemented by simplifying procedures, deployment and management of data centres infrastructure, enabling more effective association, storage, filing and processing of large data sets and resources collections.
- the principle of *standardization and interoperability* of learning data and resources is based on the standardization of tools and procedures of cloud services and resources supply to provide more transparent and understandable ways to design and deploy the components for educational purposes, their presentation and incorporation into the learning environment basing on cloud-based models.
- the principles of *security and reliability* of learning services supply means that within the cloud-based infrastructure the availability and reliability (continuity) of educational services supply increases to provide more stable performance in the environment, getting the right amounts of necessary resources and data in time to avoid or reduce the threat of data loss, unauthorized access, etc.
- the principle of *innovation* is realized with the ability to order and pay for the cloud-based services delivery as soon as they are used, to provide the freedom of choice and experimentation with different types of electronic resources, software, computer platforms and technologies in the learning process, expanding the share of investigative approach in learning, contributes to the development of skills for collaboration in the learning process, joint research and data analysis processes (Table 1).

**Table 1.** The principles of the cloud-based learning and research environment formation

| Open education | Open science | Specific to cloud-based systems |
|---|---|---|
| Mobility of students and teachers; equal access to educational systems; qualitative education; the structure and implementation of educational services formation | Technologies use; open access; research collaboration; interaction with society; innovative nature | Adaptability; personification of services delivery; ICT infrastructure unification; full-scale interactivity; flexibility and scalability; data and resources consolidation; standardization and interoperability of learning data and resources; safety and reliability; innovation |

The formation of the cloud-based learning and research environment of higher education institution should be based on the principles of open education, open science, and specific principles, in particular, adaptability; personification of services delivery; ICT infrastructure unification; full-scale interactivity; flexibility and scalability; data and resources consolidation; standardization and interoperability of learning data and resources; safety and reliability; innovation and others.

Taking into account these principles, as well as the features of the structure and the use of the cloud-based information-analytical network tools at the learning and research environment design it will be possible to facilitate the expansion of access to high-quality and large-scale information resources, to a wide range of information services offered by the learning and research information networks, practically unlimited range of users regardless of their age, gender, citizenship, location, etc.

## 3 Implementation

The cloud-based learning and research environment is the system with open structure and it may be flexibly adapted to the current needs and tasks of educational process. So the general model of the cloud-based learning and research environment formation and development [17] presupposes the possibility to create different implementations of this model for certain purposes of the environment formation in the context of current tasks of different pedagogical systems. This is achieved by changing and adapting the services and methodical systems of the initial model. Due to this the general model of CBLRE formation was elaborated and implemented in the educational process of National University of Life and Environmental Sciences of Ukraine in 2015-2017.

### 3.1 The General Model of the Cloud-based Learning and Research Environment Formation

The target component of the formation model of the cloud-oriented educational-scientific environment (CBLRE) of a higher educational institution is the improvement of the quality of educational and scientific activity of the participants in the educational process of the higher education institution (HEI), which is achieved through the introduction of resources and services of CBLRE and modern methods and forms of organization of educational and scientific activities in accordance with the principles of open education and open science.

CBLRE should be formed through a lens of the functions, the implementation of which will provide a more effective organization of the educational process at the university. The CBLRE provides the following functions of the HEI's educational activities organization: the possibility of creating e-resources for training; training portal management; communications and collaborative work of participants in the educational process; access to remote tools for doing laboratory works; access to open online training resources. For the organization of scientific activities, the CBLRE functional should include: collecting and providing open access to the results of scientific research; access to databases of scientific research results; use of new research methods (big data, remote tools); application of research collaboration tools in the course of research.

The defined functional of the environment gives an opportunity to get an idea of the components of the methodical system for the organization of educational and scientific activities. In particular, the content of training will be formed on the basis of e-resources use which can be developed either by university professors based on their own software and technology infrastructure, or selected from among the open online massive courses. These are e-learning resources for the curriculum of specialist training that can be hosted on platforms of mass open online courses, e-libraries, e-encyclopedias, scientometric databases, virtual scientific communities, open research results, etc.

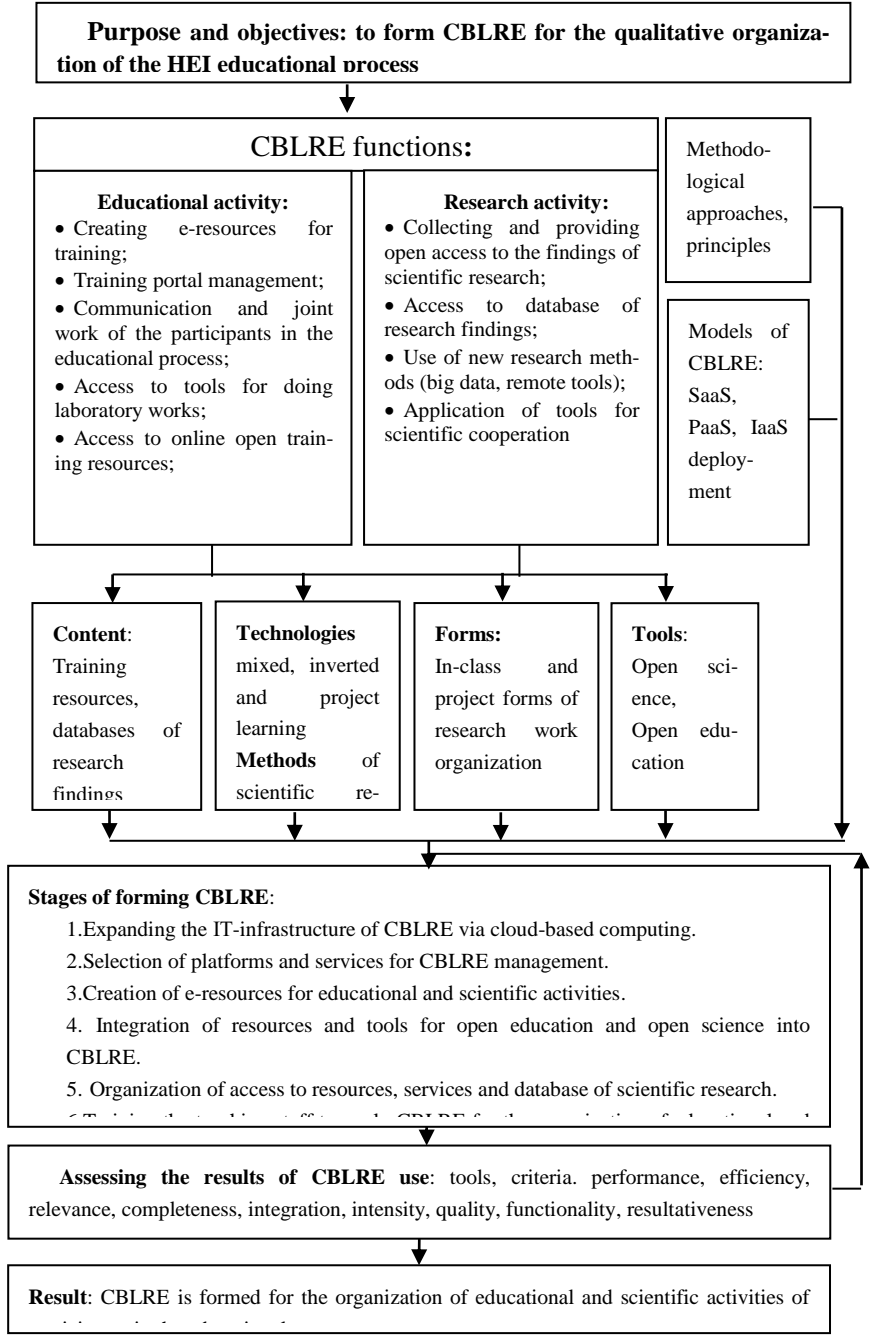

**Fig. 1.** The general model of formation the cloud-based learning and research environment of university.

Due to the availability and use of CBLRE, learners have the opportunity to change learning technologies. In particular, the technology of mixed, inverted and project learning should be actively used. A large number of on-line resources and services can be offered to students in the study of academic disciplines for self-study. Instead, in the classroom the student gets the opportunity of interactive work with the educator and students, which is reflected in the joint problem solving, tasks, discussion of topical issues, counseling, and coaching. The students are given the opportunity to build their own learning trajectory based on the experience of using the resources and services offered within the proposed learning technology. Both students and HEI educators have access to the results of scientific research open within CBLRE, which increases the efficiency of collaboration in virtual scientific communities and common sharing of data. They have an opportunity to work on cloud platforms, where large data arrays and tools for processing them are concentrated.

Thus, the application of new learning technologies and research methods based on cloud technologies, the combination of organizational forms of in-class and out-of-class activities influences the CBLRE formation, in particular, the selection of platforms and services which should form a part of the environment. We distinguish the following most effective available tools of open education: educational portals with e-learning courses to support curriculum disciplines, electronic libraries, institutional repositories with full-text resources for organizing students learning activities, as well as services for communication and collaboration. To expand the range of educational resources of manufacturing companies, HEI actively collaborates with employers and technology makers. The CBLRE of the educational institution integrates academic online courses for producers. A large number of cloud-based tools for computing, programming, modeling, designing, etc., are used to organize students' practical work. The following tools of open science should be distinguished: open conferencing systems, open journals that can be deployed inside CBLRE and used to publish the results of their own research. In addition, the important task of an educational institution is to provide access from CBLRE to scientometric databases, virtual research environments, big data processing services, etc.

6 stages of CBLRE formation are distinguished in the model. The chosen cloud deployment model (SaaS, PaaS, IaaS) will directly affect the first stage of the formation of CBLRE – the design and deployment of the HEI IT infrastructure in accordance with the functional environment. The server platforms are expanded, the campus network is configured, the access to Internet is provided.

At the second stage, the selection of platforms and services for managing CBLRE is carried out, among which is the database user's management platform, virtualization management and server capacities clustering, content management in education, etc.

At the third stage, the content base of the environment is formed, namely, e-resources for educational and scientific activities are created in accordance with the curricula and future specialists training programs.

The fourth stage is the integration of resources and tools of open education and open science into CBLRE. At this stage, the analysis and selection of cloud resources and services for use in educational and scientific activities is carried out. We deter-mine the mechanisms of their integration into academic disciplines and scientific researches. These may be open online courses, platforms for doing laboratory works, big data processing tools, spatial data processing services, etc.

At the fifth stage, access to the database of scientific research, scientometric databases, and tools for joint research is provided.

At the sixth stage, it is necessary to ensure training of the academic staff to use CBLRE for the organization of educational and scientific activities, to familiarize them with the methods and forms of training and research organization.

Criteria and tools should be developed to assess the results of applying the formed CBLRE in the HEI. To assess the infrastructure component of CBLRE, it is advisable to apply the performance, efficiency and availability criteria. To assess the resource component, we need to apply the criteria of relevance, completeness, integration into the training courses, intensity of use, quality of resources and functionality of services. The resultative component may be assessed by measuring the level of satisfaction of the participants in the educational process and the students' progress.

The established model of CBLRE formation of educational and scientific activities organization at HEI has its own features in selecting the appropriate resources and services for different specialties. The peculiarities of CBLRE formation for training future IT specialists lie in the fact that the availability of a large number of modern information technologies and resources, which may be used for the professional development of students of IT specialties, prompts the educators to change the teaching methods, to switch to more effective methods of teaching students, using new information technologies, multimedia, cloud services, etc.

The use of specially created learning resources, such as video tutorials, video lectures, lessons in e-learning courses and access to academic resources of Microsoft, Cisco, IBM, Intel, etc., allows students to improve their individual work performance on learning material and their satisfaction with the learning process. Much of scientific literature in the field of Computer Science is placed in various scientometric databases, therefore, it is necessary to ensure educators' and students' access to the Web of Science or Scopus platforms.

### 3.2 The Results of Experimental Work

The experimental research on the CBLRE formation was carried out at the National University of Life and Environmental Sciences of Ukraine (engaging 388 students). In order to evaluate CBLRE by the resultative criterion, the dynamics of the students' progress results was studied when CBLRE components were being added on a phased basis. At the first stage (Fig. 2), the components of the environment were not used to organize the learning process (NULL); at the second stage (M) CLMS Moodle was implemented and electronic training courses were developed; at the third stage, CBLRE was complemented by academic resources of Microsoft, Cisco, IBM, integrated into e-learning courses and were used by students for self-study. At the fourth stage, platforms were introduced for practical tasks, programming, data processing, open access to scientific information databases. All of these resources and services were available from CBLRE and used for practical tasks.

Since academic years served as the original sample during the experiment, in which, as it was noted, the used components varied, it is expedient to use them as a factor. In particular, this will be interesting due to the fact that for the last two years, all components of the university environment have been used. And as it was proved by the preliminary graphical analysis, the latter has a significant impact on the level of student training.

For comparison, we will construct a histogram, where academic years will serve as clusters – stages of implementing the corresponding components in CBLRE (Fig. 2).

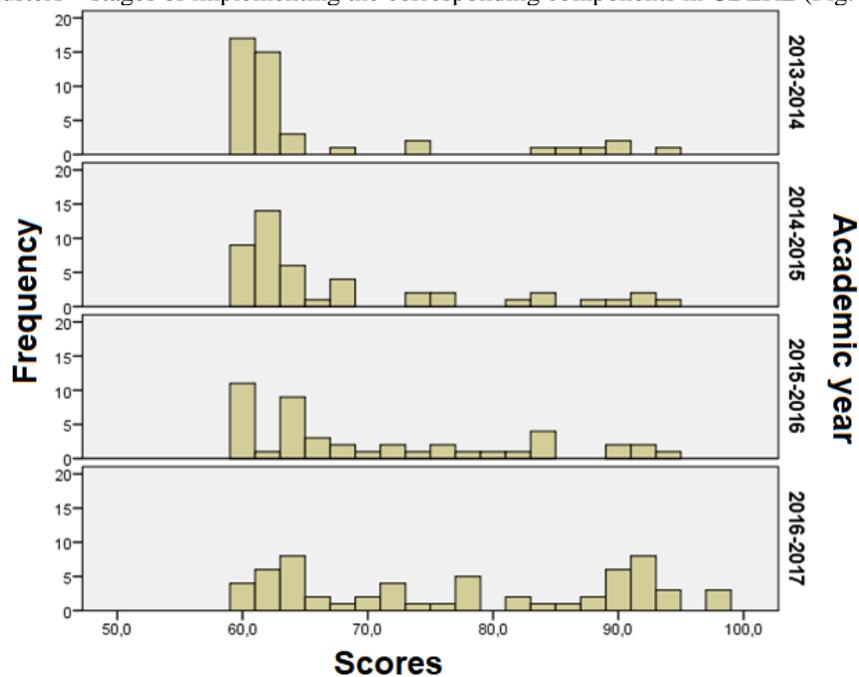

**Fig. 2.** Histogram of academic record results by years

As we can see from Fig. 2, in the academic year 2016-2017 there was a change in the frequency of the received grades: we can observe an increase of high grades according to academic record, indicating the effectiveness of the use of all three components of the environment.

However, graphical analysis alone is not sufficient for obtaining reliable statistical estimates. This requires the use of the necessary statistical tools. Since independent and dependent variables relate to different measurement scales of regression analysis, it is not appropriate. If the set of independent variables includes categorical variables, the methodology of communications estimation is shifted to the study of intergroup differences, which requires the use of ANOVA method analysis of variance. And considering that the factor in our case is one (academic year), we use a single factor analysis of variance.

The analysis of variance allows assessing the sampling for homogeneity, that is, the similarity (equality). By applying a single factor analysis of variance, the mean values of each sample are compared among themselves and the overall level of significance of the differences is calculated. We construct a table of descriptive statistics (Table 2). As we see from this table, the mean values of grades are increasing annually. However, when we evaluate the confidence interval of the mean, we see a significant corridor. This indicates a significant variation for the true mean.

**Table 2.** Descriptive statistics

| Academic year | No. | Average grade | Standard deviation | Standard error | 95% confidence interval for the mean | | Minimum | Maximum |
|---|---|---|---|---|---|---|---|---|
| | | | | | Lower limit | Upper limit | | |
| 2013-2014 | 88 | 65,500 | 9,5905 | 1,0223 | 63,468 | 67,532 | 60,0 | 93,0 |
| 2014-2015 | 92 | 67,804 | 10,1193 | 1,0550 | 65,709 | 69,900 | 60,0 | 93,0 |
| 2015-2016 | 88 | 70,182 | 10,6615 | 1,1365 | 67,923 | 72,441 | 60,0 | 93,0 |
| 2016-2017 | 120 | 77,083 | 12,8590 | 1,1739 | 74,759 | 79,408 | 60,0 | 98,0 |
| Total | 388 | 70,691 | 11,0499 | ,6049 | 69,588 | 71,880 | 60,0 | 98,0 |

The table of obtained single factor analysis of variance (Table 3) for experimental data gives us the value of the F-criterion (Fischer's criterion) equal to 22.015. The corresponding critical value of the F-criterion for the intergroup degree of freedom (4-1) = 3 and the intra-group (388-4) = 384 at α = 0.05 is 2.628. As we see, the Fisher's criterion calculated according to experimental data exceeds the critical one – 22,015> 2,628. Consequently, the null hypothesis of the analysis of variance is rejected – the absence of the effect of factor under consideration. Accordingly, based on the results of analysis of variance we can state that the use of CBLRE affects the learning outcomes. And we can state that the academic grade average has grown by 11.5%.

**Table 3.** Single Factor Analysis of Variance

| | Sum of squares | Degrees of freedom | Mean-square value | F | Value |
|---|---|---|---|---|---|
| Between the groups | 8064,151 | 3 | 2688,050 | 22,015 | 0,000 |
| Inside groups | 46886,736 | 384 | 122,101 | | |
| Total | 54950,887 | 387 | | | |

In the long term, it is necessary to investigate the effectiveness of CBLRE by resource and infrastructure criteria. To do this, appropriate indicators and an assessment scale should be developed.

## 4     Conclusions

The general model of the university cloud-based learning and research environment formation proved to be a reasonable framework to deliver and research the cloud-based learning resources and components. The ways of methods selection based on the proposed model and the prospects for their use within the learning systems of higher education were considered. The integration of resources and services into a single, cloud-oriented educational environment contributes not only to improving the efficiency of access to the necessary tools, it enables the use of advanced training technologies, big data processing tools, and other sources of open education and science. The use of cloud-based technologies in building an IT-infrastructure ensures CBLRE performance and efficiency. The experiment proved that by applying CBLRE resources and services the results of students' training increase by more than 11%.